\documentclass{aa}

\usepackage{graphicx}
\usepackage{tabularx} 
\usepackage{titlesec}
\newcounter{subsubsubsection}[subsubsection]

\usepackage{txfonts}

\usepackage{hyperref}
\hypersetup{colorlinks=true,linkcolor=blue,citecolor=blue,filecolor=blue,urlcolor=blue}

\begin{document}

   \title{A search for optical counterparts in quiescent black hole X-ray transients}

   \author{I. V. Yanes-Rizo \inst{1,2}\fnmsep\thanks{\email{idaira.yanes@gmail.com} \\ $\dagger$ Deceased}
          \and
          J. Casares \inst{1,2}
          \and
          M. A. P. Torres \inst{1,2}
          \and
          V. S. Dhillon \inst{3,1}
          \and
          T. R. Marsh \inst{4} $\dagger$
          \and
          M. Armas Padilla \inst{1,2}
          \and
          P. G. Jonker \inst{5}
          \and
          T. Muñoz-Darias \inst{1,2}
          \and
          S. Navarro Umpiérrez \inst{1,2}
          \and
          D. Steeghs \inst{6}
          }

   \institute{Instituto de Astrofísica de Canarias, E-38205 La Laguna, S/C de Tenerife, Spain\\
         \and
             Departamento de Astrofísica, Universidad de La Laguna, E-38206 La Laguna, S/C de Tenerife, Spain\\
         \and
             Astrophysics Research Cluster, School of Mathematical and Physical Sciences, University of Sheffield, Sheffield S3 7RH, UK\\
         \and
             Department of Physics, University of Cambridge, Cambridge, CB3 0HA, UK\\
         \and
             Department of Astrophysics/IMAPP, Radboud University, P.O. Box 9010, 6500 GL Nijmegen, the Netherlands\\
         \and
             Department of Physics, University of Warwick, Gibbet Hill Road, Coventry CV4 7AL, UK\\
             }

   \date{zzz}

  \abstract
   {Dynamical mass measurements of compact stars in X-ray transients demand the detection of optical/near infrared counterparts in quiescence. Out of the 73 black-hole candidates in X-ray transients, optical and near-infrared quiescent counterparts have only been identified for 34 objects. We present ULTRACAM photometric observations of nine candidate black hole X-ray transients with no reported counterparts in quiescence, complemented with data from the public surveys DECaPS and Pan-STARRS. In addition, we analyze photometry of three sources (SWIFT J1539.2-6227, XTE J1817-330 and XTE J1818-245) obtained during their discovery outburst. The data provide the first optical identifications and precise astrometry of four targets (MAXI J1348-630, SWIFT J1539.2-6227, XTE J1726-476 and XTE J1817-330) plus $3\sigma$\,lower limits to the quiescent optical magnitudes for an additional five (MAXI J0637-430, 4U 1755-338, MAXI J1803-298, XTE J1818-245 and MAXI J1828-249). Of these five, 4U 1755-338 was found to be active during our ULTRACAM observations and we use our images to derive refined astrometric coordinates. We use the photometric magnitudes and colors to place preliminary constraints on the orbital periods and spectral types of the companion stars. Finding charts of all the targets are also provided to facilitate future follow-up studies. Finally, we present updated astrometry for XTE J1650-500 using archival FORS2 images.}


   \maketitle

\section{Introduction}
Galactic stellar-mass black holes (BHs) have been mainly identified in X-ray transients (XRTs), a sub-class of X-ray binaries that exhibit dramatic X-ray outbursts with typical timescales of decades-centuries \citep{mcclintock2006}. To date, 73 XRTs hosting BH candidates have been unveiled over the past $\sim 60$\,years \citep[see][]{corral2016}, and yet only 20 have been dynamically confirmed through radial velocity studies. This disparity arises largely from the challenges associated with measuring the spectra of companion stars at very faint quiescent magnitudes, which are essential for precise mass determinations \citep{casares2014}. This limitation is due to their location in the Galactic plane, where interstellar extinction is greater. Consequently, our understanding of the fundamental parameters of these systems and thus, of the broader picture of BH formation in XRTs, is limited by small number statistics and selection biases \citep[e.g.][]{jonker2021}. In this work, we aim to expand the known sample of quiescent BH XRT detections by conducting a systematic photometric search for candidate systems among historic and newly discovered XRTs. This effort represents a stepping stone toward more comprehensive demographic studies, allowing us to identify and prioritize promising targets for future dynamical mass measurements of BHs.

\section{Sample selection}
\label{sec:sampleselection}
The online version of the BlackCAT catalogue\footnote{\url{https://www.astro.puc.cl/BlackCAT/transients.php}} contains 73 BH transients detected until 2025. Out of these, confident optical or near infrared (NIR) counterparts exist for 56 sources in outburst and only 34 in quiescence. In addition, five have bright interlopers at $\leq 1$\,arcsec (XTE J2012+381, XTE J1652-453, MAXI J1836-194, SWIFT J1658-4242 and MAXI J1631-479) while ten are heavily reddened and do not posses good astrometric positions due to the lack of a known counterpart, even in outburst (Cen X-2, GRS 1730-312, XTE J1755-324, GRS 1737-31, XTE J1748-288, EXO 1846-031, GS 1734-275, SAX J1711.6-3808, MAXI J1810-222 and MAXI J1728-360, with the former four also having large X-ray errors of $\geq$\,1 arcmin). Among the 34 sources with a known quiescent counterpart, 20 have already been confirmed as dynamical BHs, with another six having indirect evidence based on H$\alpha$\,scaling relations \citep[][]{mata2015, torres2021, casares2023, yanes2024, yanes2025, corral2025}. 

For the 28 candidates with accurate astrometry but no quiescent counterparts, we have estimated quiescent $r$-band magnitudes either based on typical outburst amplitude of 6 mag \citep[i.e. the peak in the observed distribution;][]{shahbaz1998, lopez2019} or from pre-existing NIR quiescent magnitudes, all obtained from BlackCAT. In the latter case, we have adopted the colors of a commonly observed K5 donor star \citep{casares2025} and $E(B-V)$\, reddening information, also from the BlackCAT catalogue.

According to \citet{shahbaz1998}, the outburst amplitude appears to be anticorrelated with orbital period for periods $\leq 1$\,d (i.e. for systems with companion stars close to the main sequence), with values ranging between $\sim 4-8$\,mag. For the case of evolved stars in longer period systems, such as XTE J1550-560 or V404 Cyg, the amplitudes range between $4-6$ mags. Thus, we estimate that using this relation to infer the magnitude of the quiescent counterpart once the orbital period is known would result in errors of up to $\pm 2$\,mag. On the other hand, disregarding the reddening error (which varies with each system), the uncertainty caused by adopting K5 V colors is more modest, at $\pm 0.5$\,mags for spectral types K0-M0. Following this procedure, we have selected nine targets with predicted $r$-band quiescent magnitudes in the range $\sim 21 - 24$\, and astrometric accuracy $< 1$\,arcsec, which are therefore amenable for identification with deep optical imaging. These are listed in Table~\ref{tab:coords} with their coordinates and associated positional uncertainties as reported in the literature. Furthermore, we provide a refined astrometric position for the dynamically confirmed BH XTE J1650-500 based on archival images (see Sect.~\ref{sec:xtej1650}).

\section{Optical observations and data reduction}
\subsection{NTT time-resolved photometry}
Time-resolved photometry was carried out using the triple arm ULTRACAM camera \citep{dhillon2007} mounted on the 3.5-m New Technology Telescope (NTT) in La Silla observatory, Chile. The ULTRACAM design features three 1024 × 1024 pixel frame-transfer CCDs that together cover a 5x5 arcmin field of view, with a plate scale of 0.3 arcsec per pixel. Only for observations of MAXI J1348 on the night of 15 March 2021 we apply a binning of 2x2 due to the bad seeing conditions. This configuration permits the simultaneous acquisition of images in the super SDSS filters $u_s$, $g_s$, and $r_s$ bands \citep{dhillon2021}, which is especially advantageous for monitoring rapid variability across multiple wavelengths. Our strategy was to obtain deep images ($r \sim 25 - 26$\, in one hour integration) from a series of short 20 s, sky-limited exposures, with only 24 msec dead-time, taken under relatively good seeing conditions of up to 1.2 arcsec. Since the ULTRACAM detectors have low read out noise, deep images can be built up without a significant signal-to-noise ratio (SNR) penalty compared to a single long-exposure image, while also capturing variability information on the target.

Observations were conducted during two separate campaigns, 15–16 March 2021 and 23–24 March 2023, where the seeing was highly variable, ranging between $0.8 - 2.8$\,arcsec. A log of the observations, including exposure times and observing conditions, is provided in Table~\ref{tab:log}. Standard data reduction procedures were applied to the raw images. Bias subtraction and flat field correction were performed using the HiPERCAM\footnote{\url{https://github.com/HiPERCAM/hipercam}} pipeline. This was also employed to perform aperture photometry of isolated stars that were fluxed calibrated with the Dark Energy Camera Plane Survey 2 \citep[DECaPS2;][]{saydjari2023} or the Pan-STARRS Data Release 2 catalogue \citep{chambers2016}. The field of MAXI J0637-430 was not covered by these surveys. In this case, the photometry was calibrated with photometric zero points established from individual images of MAXI J1348-630 taken on 23 March 2023 under the same transparency and seeing conditions. Due to the extreme weakness of all the targets in the $u_s$ band, we performed the analysis only for the $g_s$ and $r_s$-band images.

\subsection{VLT and Magellan imaging}
We also analyzed optical images of XTE J1650-500, SWIFT J1539.2-6227 and XTE J1817-330 taken at different phases of their discovery outbursts. This allow us to derive improved coordinates and/or outburst apparent magnitudes.

The optical counterpart to XTE J1650-500 is clearly detected in a 5 s $R_{special}$-band image obtained on 10 June 2002 with the FOcal Reducer and low-dispersion Spectrograph 2 \citep[FORS2;][]{appenzeller1998} which was mounted on the 8.2-m Unit 1 Very Large Telescope at Paranal observatory, Chile. The image was taken for the acquisition of the near-quiescence spectroscopy presented in \citet{sanchez2002} and \citet{orosz2004}. The XRT was recorded under 0.7 arcsec seeing; sampled with a 0.252\,arcsec per pixel plate scale. XTE J1817-330 was targeted on the night of 10 January 2006, at the start of its discovery outburst, with the Low Dispersion Survey Spectrograph (LDSS3) attached to the 6.5-m Magellan Clay telescope at Las Campanas observatory, Chile. A series of 5-10 s $g$-band images were taken under 1.2 arcsec seeing sampled with a 0.188\,arcsec\ pixel$^{-1}$ scale. SWIFT J1539.2-6227 was observed on 10 February 2009, $\sim 2.5$ months after the outburst onset, with the Inamori-Magellan Areal Camera and Spectrograph \citep[IMACS;][]{dressler2011} mounted on the 6.5-m Magellan Baade telescope. Five 10 s $g$-band images were obtained under 1.1 arcsec seeing sampled with a 0.199\,arcsec\ pixel$^{-1}$ plate scale.

The LDSS3 and IMACS $g$-band images were corrected from bias and flat-fielded using standard routines running within {\sc pyraf}\footnote{\url{https://github.com/iraf-community/pyraf}}. Aperture photometry of the targets and field stars (flux calibrated in DECaPS2) was performed with the HiPERCAM pipeline and differential photometry used to obtain the target apparent magnitude. Note here that the field stars selected for the flux calibrations have negligible error contribution to the apparent magnitudes of any of the optical counterparts identified in this paper.

\subsection{Image selection by seeing}
\label{sec:seeingselection}
In order to identify faint optical counterparts in the ULTRACAM data, it is crucial to obtain a combined image with the highest SNR possible. To achieve this, we implemented a filtering criterion based on the atmospheric seeing conditions during our observations. Clearly the maximum SNR and depth that can be reached by co-adding images under variable seeing will be a trade off between seeing cut-off and the number of images selected. This filtering process was implemented as follows: we first measured the full width at half maximum (FWHM) of several point-like sources in each image using the HiPERCAM pipeline to provide a robust estimate of the seeing in each exposure. We then produced co-added images for a set of cut-off seeing values between $0.9$\,arcsec and $2$\,arcsec in steps of $0.05$\,arcsec and performed photometry on each image across the entire field of view. The limiting magnitude was subsequently obtained by determining the magnitudes of objects with signals three times above the background noise. Following this procedure, we identified the optimal seeing cut-off that produces the deepest image, and these are the images that were used in our subsequent photometric analysis. 

\begin{figure}
	\centering \includegraphics[width=\columnwidth]{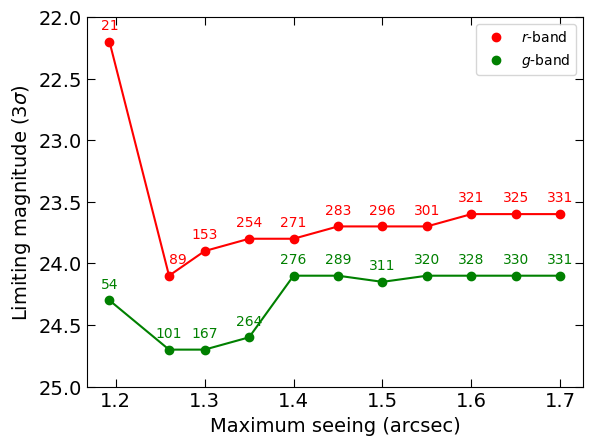}
    \caption{Limiting magnitudes as a function of the seeing cut-off for the X-ray transient SWIFT J1539.2-6227. The red and green dots represent the values obtained for the $r$ and $g$-band, respectively. The red and green numbers indicate the number of combined frames. The deepest magnitude is obtained by selecting images with a seeing up to 1.25 arcsec.}
    \label{fig:seeingselection}
\end{figure}

Figure~\ref{fig:seeingselection} shows an example of how the limiting magnitudes vary as a function of the cut-off seeing for the XRT SWIFT J1539.2-6227. In this instance, restricting the data to images with seeing up to 1.25 arcsec yielded the deepest limiting magnitude, thereby maximizing our chance to detect the faint counterpart. For information, in Table~\ref{tab:log} we provide the cut-off seeing and the effective exposure time achieved for each individual object. 

In addition, we searched for detections of the quiescent optical counterparts in the DECaPS2 and Pan-STARRS multi-band catalogues and images. When suitable, we performed photometry on the individual images to derive the magnitudes of the optical counterparts or a $3 \sigma$ lower limit when these are not detected.

\subsection{Astrometry}
In order to accurately pinpoint the positions of faint optical counterparts (especially those located in crowded stellar fields), we conducted an astrometric calibration using the Gaia\footnote{\url{https://github.com/Starlink/starlink/tree/master/applications/gaia}} image tool. This cross references our images with stars from the Gaia Data Release 2 catalog \citep{gaia2018}, fitting the celestial positions with a least squares minimization procedure. This calibration yielded astrometric solutions with rms residuals ranging from 0.02 to 0.08 arcsec. Such accuracy is crucial not only for the confident identification of faint optical counterparts but also for facilitating reliable cross matching with multi wavelength datasets in follow up studies.

\subsection{PSF photometry}
\label{sec:psfphotometry}
The presence of interloper stars in crowded fields, for example in the cases of XTE J1817-330 and XTE J1818-245, can significantly complicate the identification of the optical counterparts. To mitigate this issue, we employed point spread function (PSF) photometry using the DAOPHOT package \citep{stetson1987} implemented in {\sc pyraf}, which is well-suited for crowded field analysis. In our approach, we modeled the PSF of stars using a Moffat distribution. This model was applied specifically to a selection of stars that were both isolated and unsaturated, ensuring that the derived PSF was reliable. We then fitted the resulting Moffat PSF model to the stars in the vicinity of each target. For each individual field and filter, we adjusted the fitting and PSF radii to optimize the match between the model and the observed stellar profiles, thereby minimizing residuals. This fine-tuning process was essential to accurately subtract the light from contaminating interloper stars and to reveal the underlying faint optical counterparts. When recovered in the residual images, the counterparts were included in the photometry and the PSF fit was repeated to derive their magnitudes.

\begin{figure*}
	\centering \includegraphics[width=\textwidth,height=\textheight,keepaspectratio]{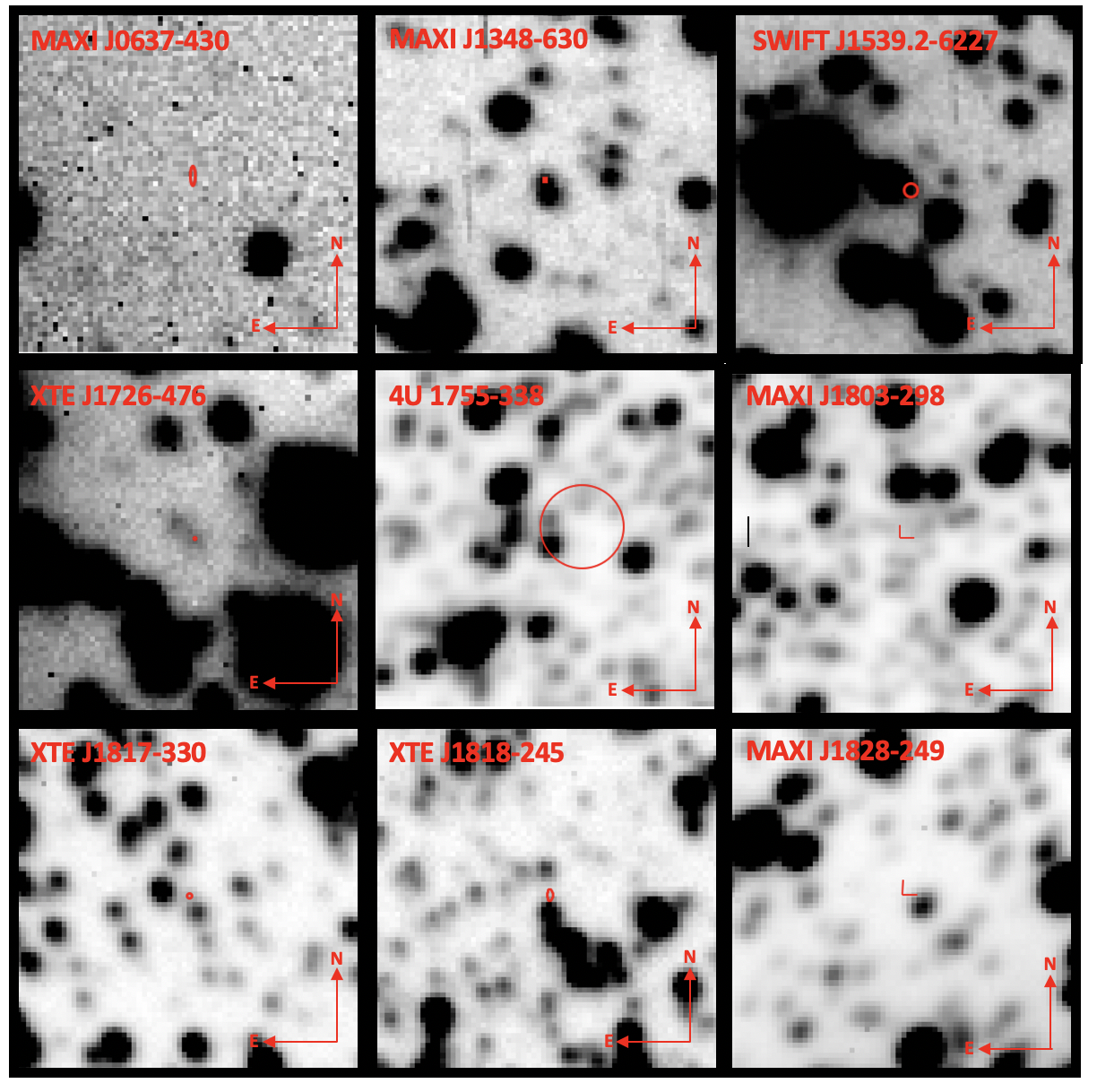}
    \caption{Deepest $r$-band images for the sample of the nine BH candidates. The location of the objects are indicated by red ellipses, taking into account the positional errors reported in the literature (see Table~\ref{tab:coords}). For the sake of visibility, we mark the position of J1803 and J1828 with red tick marks instead of ellipses, as the positional errors are too small to be resolved. The field of view is 20 x 20 arcsec each.}
    \label{fig:finders}
\end{figure*}

\section{Methodology}
\label{sec:extinctionlaws}
In this section we present the methods that will be used in Sect.~\ref{results_discussion} to constrain the orbital period (i.e. the most fundamental binary parameter) and the spectral type  of the companion star. In Sect.~\ref{final_considerations} we will detail the systematics that can affect our calculations.

For the cases in which the optical counterpart is detected, we first correct the observed magnitudes for interstellar extinction by applying the extinction relationships established by \citet{schlafly2011}: $A_V = 3.1\, E(B-V)$\,; $A_r = 2.285\, E(B-V)$\,; $A_g = 3.303\, E(B-V)$\,; $A_i = 1.698\, E(B-V)$\,; $A_z = 1.263\, E(B-V)$. These relations allow us to transform $E(B-V)$ into absolute extinction values ($A_X$) in the corresponding photometric bands. By subtracting these extinction values from the observed magnitudes, we obtain the intrinsic colors of the sources. For objects for which the hydrogen column density ($N_{\mathrm{H}}$) has been measured from X-ray fitting, we employ the relation of \citet{guver2009}, $N_{\mathrm{H}} \;(\mathrm{cm}^{-2}) = (2.21 \pm 0.09) \times 10^{21} A_V$. Otherwise, color excess $E(B-V)$ values were taken from the extinction maps of \citet{schlafly2011}. They correspond to the total reddening along the line of sight to the object. Once these parameters are determined, we can constrain the spectral type of the companion star by comparing the dereddened color to standard color indices \citep[Table 3 in][]{covey2007}. This constraint is an upper limit to the spectral type because E(B-V) may contain background extinction, $N_{\mathrm{H}}$\,can have a contribution local to the XRT \citep{jonker2004} and/or the light from the companion star will likely be diluted by a (blue) flux contribution from the residual accretion disc. Previous work has shown that the intrinsic color obtained from simultaneous or contiguous photometry provides a reasonable estimate of the spectral type of the companion or robust upper limit in cases where there is significant disc contamination \citep{casares2023, yanes2024}.

As already introduced in Sect.~\ref{sec:sampleselection}, the $V$-band amplitude ($\Delta V$) of an outburst in XRTs is thought to be correlated with the orbital period ($P_{\mathrm{orb}}$) through the expression $\Delta V = 14.36 - 7.63\,log(P_{\mathrm{orb}} (h))$ \citep{shahbaz1998}. However, it is important to note that this relation is based on a limited sample of objects and outbursts \citep[$\sim 8$; see][]{lopez2019}. In addition, the orbital inclination angle can significantly influence the observed outburst amplitude, as reported by \citet{miller2011}. Keeping in mind these considerations, we will use the $\Delta V - P_{\mathrm{orb}}$ relation to obtain rough estimates for the orbital period by assuming that the Sloan $g$ magnitude can be equated to Johnson V. In many cases, the outburst amplitudes will be lower limits because the peak magnitude was missed, the source is not detected in quiescence, or both.

The mean stellar density $\bar{\rho}$ for a Roche-lobe filling star can be obtained from $P_{\mathrm{orb}}$ through $\bar{\rho} \approx 110 \times P^{-2}_{\rm{orb}}$\,gr cm$^{-3}$ \citep{frank2002}. Thus, from the constraint on $P_{\mathrm{orb}}$ we will bound the spectral type of the companion star by comparing its $\bar{\rho}$ with that of main sequence stars. Conversely, the $\bar{\rho} - P_{\mathrm{orb}}$ relation will be used to set limits on $P_{\mathrm{orb}}$ from the constraints on the spectral type. For this, we computed the density values using the stellar mass and radius reported online by \citet{pecaut2013}. We stress that, in our range of interest $P_{\mathrm{orb}} \lesssim 1$\,d, adopting a main sequence model is a reasonable and rather conservative assumption (see Sect.~\ref{final_considerations}). 

\begin{table*}
    \centering
    \caption{Astrometric positions of the observed objects.}
    \label{tab:coords}
    \begin{tabular}{ccc|ccc}
        \hline
        Object & \multicolumn{2}{c}{Coordinates from literature} & \multicolumn{2}{c}{Coordinates from this work} & Ref. to astrometry \\ 
        \cline{2-3} \cline{4-5}
               & RA (J2000) & DEC (J2000) & RA (J2000) & DEC (J2000) & \\ \hline
        MAXI J0637-430 & \begin{tabular}{c} 06:36:23.7 \\ $(\pm 0.2")$ \end{tabular} & \begin{tabular}{c} -42:52:04.1 \\ $(\pm 0.7")$ \end{tabular} & \begin{tabular}{c} $--$ \end{tabular} & \begin{tabular}{c} $--$ \end{tabular} & [1] \\ \hline
        MAXI J1348-630 & \begin{tabular}{c} 13:48:12.79 \\ $(\pm 0.03")$ \end{tabular} & \begin{tabular}{c} -63:16:28.48 \\ $(\pm 0.04")$ \end{tabular} & \begin{tabular}{c} $--$ \end{tabular} & \begin{tabular}{c} $--$ \end{tabular} & [2] \\ \hline
        SWIFT J1539.2-6227 & \begin{tabular}{c} 15:39:11.96 \\ $(\pm 0.5")$ \end{tabular} & \begin{tabular}{c} -62:28:02.30 \\ $(\pm 0.5")$ \end{tabular} & \begin{tabular}{c} 15:39:11.92 \\ $(\pm 0.07")$ \end{tabular} & \begin{tabular}{c} -62:28:02.37 \\ $(\pm 0.07")$ \end{tabular} & [3] \\ \hline
        XTE J1650-500 & \begin{tabular}{c} 16:50:00.92 \\ $(\pm 0.31")$ \end{tabular} & \begin{tabular}{c} -49:57:44.1 \\ $(\pm 0.31")$ \end{tabular} & \begin{tabular}{c} 16:50:00.95 \\ $(\pm 0.05")$ \end{tabular} & \begin{tabular}{c} -49:57:44.25 \\ $(\pm 0.05")$ \end{tabular} & [4] \\ \hline
        XTE J1726-476 & \begin{tabular}{c} 17:26:49.30 \\ $(\pm 0.1")$ \end{tabular} & \begin{tabular}{c} -47:38:25.5 \\ $(\pm 0.1")$ \end{tabular} & \begin{tabular}{c} 17:26:49.31 \\ $(\pm 0.05")$ \end{tabular} & \begin{tabular}{c} -47:38:25.63 \\ $(\pm 0.05")$ \end{tabular} & [5] \\ \hline
        4U 1755-338 & \begin{tabular}{c} 17:58:39.84$^{\alpha}$ \\ $(\pm 2.8")$ \end{tabular} & \begin{tabular}{c} -33:48:27.64$^{\alpha}$ \\ $(\pm 2.8")$ \end{tabular} & \begin{tabular}{c} 17:58:40.01 \\ $(\pm 0.04")$ \end{tabular} & \begin{tabular}{c} -33:48:28.80 \\ $(\pm 0.04")$ \end{tabular} & [6] \\ \hline
        MAXI J1803-298 & \begin{tabular}{c} 18:03:02.79178 \\ $(\pm 0.00003")$ \end{tabular} & \begin{tabular}{c} -29:49:49.41220 \\ $(\pm 0.00007")$ \end{tabular} & \begin{tabular}{c} $--$ \end{tabular} & \begin{tabular}{c} $--$ \end{tabular} & [7] \\ \hline
        XTE J1817-330 & \begin{tabular}{c} 18:17:43.53 \\ $(\pm 0.2")$ \end{tabular} & \begin{tabular}{c} -33:01:07.57 \\ $(\pm 0.2")$ \end{tabular} & \begin{tabular}{c} 18:17:43.51 \\ $(\pm 0.04")$ \end{tabular} & \begin{tabular}{c} -33:01:07.68 \\ $(\pm 0.04")$ \end{tabular} & [8] \\ \hline
        XTE J1818-245 & \begin{tabular}{c} 18:18:24.43 \\ $(\pm 0.2")$ \end{tabular} & \begin{tabular}{c} -24:32:17.96 \\ $(\pm 0.4")$ \end{tabular} & \begin{tabular}{c} 18:18:24.43 \\ $(\pm 0.08")$ \end{tabular} & \begin{tabular}{c} -24:32:18.08 \\ $(\pm 0.08")$ \end{tabular} & [9] \\ \hline
        MAXI J1828-249 & \begin{tabular}{c} 18:28:58.0748 \\ $(\pm 0.0004")$ \end{tabular} & \begin{tabular}{c} -25:01:45.7276 \\ $(\pm 0.0006")$ \end{tabular} & \begin{tabular}{c} $--$ \end{tabular} & \begin{tabular}{c} $--$ \end{tabular} & [10] \\ \hline
    \end{tabular}
    \tablefoot{$^{\alpha}$ At the time of submission, an updated position has been provided in the on-line catalogue by \citet{fortin2024} but we note that this corresponds to a star located $\sim 2$\,arcsec north of the true counterpart identified by \citet{wachter1998} and, therefore, these coordinates are not provided here. \\ 
    $[1]$ \citet{russell2019}, $[2]$ \citet{russell2019j1348}, $[3]$ \citet{krimm2011}, $[4]$ \citet{gallo2008}, $[5]$ \citet{steeghs2005}, $[6]$ \citet{waddell2026}, $[7]$ \citet{wood2023}, $[8]$ \citet{rupen2006}, $[9]$ \citet{rupen2005}, $[10]$ \citet{gaia2021}}
\end{table*}

\section{Results}
\label{results_discussion}
Figure~\ref{fig:finders} displays zoomed-in finding charts derived from the deepest $r$-band images generated as described in Sect.~\ref{sec:seeingselection} for the fields of the nine BH candidates listed in Table~\ref{tab:coords}. These finding charts not only highlight the precise positions of the candidates but also include reference stars to facilitate subsequent follow up observations. The astrometric positions of the optical counterparts are given in Table~\ref{tab:coords}. Table~\ref{tab:photometricresults} summarizes our photometric results while the implied binary constraints are provided in Tables~\ref{tab:binaryconstraints} and~\ref{tab:finalconstraints}.

\subsection{MAXI J0637-430}
MAXI J0637-30 (hereafter J0637) was discovered by the Monitor of All-sky X-ray Image (MAXI) Gas Slit Camera (GSC) on November 2019 \citep{negoro2019}. The optical outburst peaked at $g = 16.23 \pm 0.01$ (\citealt{baglio2020}; see also \citealt{hambsch2019}). The global X-ray spectral and timing properties are consistent with an accreting black hole \citep{russell2019, tomsick2019}. \citet{teratenko2021} set limits for the orbital period of 1.5 and 4 h from the detection of Hydrogen emission lines \citep{strader2019} and the fitting of the spectral energy distribution, respectively. Further studies, based on the properties of the X-ray and optical outburst spectra, proposed a tentative 2.2 h orbital period \citep{soria2022}. 

J0637 was observed on all four nights of our NTT campaigns. Fig.~\ref{fig:finders} displays the deepest image obtained following the procedure described in Sect.~\ref{sec:seeingselection}. Based on the noise characteristics of the final image, we derive $3\sigma$ quiescent limiting magnitudes of $g > 25.8$ and $r > 25.3$ \,for J0637. Despite reaching a significant depth in our observations, no counterpart was detected at the expected location. This non-detection implies that the optical emission from the source is very faint, as the line of sight experiences minimal interstellar extinction given its favorable location in the Galaxy ($b = -20.67$\,deg). 

We estimate an outburst amplitude: $\Delta V \approx \Delta g \geq 8.8$, which results in $P_{\mathrm{orb}} \leq 4.5$\,h employing the \citet{shahbaz1998} relation. Using the $\bar{\rho} - P_{\mathrm{orb}}$ relation, our constraint on the orbital period results in a mean density for the companion star > 5.4\,gr cm$^{-3}$. For a main sequence star, this limit corresponds to a spectral type later than M1 V.

\subsection{MAXI J1348-630}
MAXI J1348-630 (hereafter J1348) was discovered on January 2019 by MAXI \citep{yatabe2019}. X-ray spectral analysis and timing properties support the black hole nature of J1348 \citep{sanna2019, zhang2020}. Optical monitoring conducted by \citet{baglio2019} reported $g = 16.09 \pm 0.04$ in the outburst peak. The decay to quiescence was followed by \citet{alyazeedi2019} who provided a tentative quiescent magnitude of $i = 20.3 \pm 0.4$, obtained between two brightening episodes; the quiescent state of J1348 could not be confirmed. 

Distance estimates for J1348 have been obtained through line-of-sight HI absorption measurements with ASKAP and MeerKAT, with a most probable value of $2.2$\,kpc and an upper limit of $5.3$\,kpc \citep{chauhan2021} and, subsequently, through dust scattering ring observations that led to an estimated distance of $3.39 \pm 0.34$\,kpc \citep{lamer2021}.  Observations of its quiescent X-ray luminosity suggest that the system may have an orbital period of less than 10-20 h \citep{carotenuto2022}. 

We collected a total of 10 hours of ULTRACAM data during our NTT campaigns. The deepest image, displayed in Fig.~\ref{fig:finders}, reveals a blend of two stars separated by 1.1 arcsec, with the radio position of J1348 coincident with the fainter northern object. Figure~\ref{fig:psffinders} shows the residual image in the $r$-band after subtraction of the bright interloper using PSF photometry. We derive quiescent magnitudes of $g = 23.6 \pm 0.1$ and $r = 21.18 \pm 0.03$. In addition, we examined DECaPS2 images of the field, where J1348 and the interloper are well resolved (see Fig.~\ref{fig:decaps_finders}). We performed aperture photometry on the 24 March 2016 images and obtain $g = 23.4 \pm 0.2$\,, $r = 21.3 \pm 0.1$\,, $i = 20.5 \pm 0.1$ and $z = 19.41 \pm 0.07$. These magnitudes are consistent with our ULTRACAM photometry. Additional DECaPS2 observations in the $g$ and $r$-bands were taken in 2017 with both magnitudes remaining stable between these two epochs.

Using our quiescent $g$-band and the peak outburst magnitude, we derive an outburst amplitude of $7.5 \pm 0.1$\,mag which leads to $P_{\mathrm{orb}} \leq 7.9$\,h. Additionally, based on the mean stellar density of a Roche lobe filling main sequence star for that period ($> 1.7$\,gr cm$^{-3}$), we constrain the spectral type of the companion star to be later than G8 V. On the other hand, given $N_{\mathrm{H}} = 8.6 \times 10^{21}$\,cm$^{-2}$ \citep[][see also Sect.~\ref{final_considerations}]{tominaga2020, zhang2022, liu2022, dai2023}, we estimate $E(B-V)= 1.26 \pm 0.06$\,and derive $(g-r)_0 = 1.2 \pm 0.2$, $(r-i)_0 = 0.7 \pm 0.2$ and $(i-z) = 0.6 \pm 0.3$, which correspond to spectral types of K2, K4 and K7 V, respectively. We adopt K2 V since our ULTRACAM photometry (which provides $g$ and $r$ band photometry) are based on simultaneous observations.  

Since the distance to MAXI J1348-630 is constrained to a mean value of $\sim 2.8$\,kpc we can use the empirical relation $M_r =(4.64 \pm 0.10) - (3.69 \pm 0.16)log_{10}[P_{\mathrm{orb}}(d)]$\,\citep{casares2018} and our $r$-band detection to derive an independent constraint on the orbital period. Therefore, by bringing $r = 21.18 \pm 0.03$\,to the distance modulus equation, together with $d \sim 2.8$\,kpc and $A_r \approx 2.9$\,, we find $M_r > 6.0$\,, which leads to $P_{\mathrm{orb}} < 10.1$\,h.

\subsection{SWIFT J1539.2-6227}
\label{swift_j1539}
SWIFT J1539.2-6227 (hereafter J1539) was discovered by the Swift Burst Alert Telescope \citep{barthelmy2005} on November 2008 \citep{krimm2008} as a new Galactic transient. Observations made with the Swift UltraViolet/Optical Telescope revealed a bright source ($uvw2 = 18.07 \pm 0.03$ and $uvm2 = 17.96 \pm 0.04$) with precise coordinates that are listed in Table~\ref{tab:coords} \citep{krimm2009}. The X-ray spectral evolution and timing features \citep{krimm2011} support that J1539 is a BH candidate in a low-mass X-ray binary. \citet{lopez2019} set limits on the quiescent magnitudes of $g > 19.6$ and $r > 20.1$.

We performed astrometry and differential photometry on the outburst $g$-band images obtained in February 2009 with IMACS at the Magellan-Baade Telescope and pre-analyzed in \citet{torres2009}. We determine the position of J1539 to be RA(J2000) = 15:39:11.92 and Dec(J2000) = $-$62:28:02.37, with an uncertainty of 0.07 arcsec. This position is an improvement over the one reported by \citet{krimm2011}. Additionally, we measured a magnitude of $g = 18.25 \pm 0.01$. 

We observed J1539 in quiescence with ULTRACAM on the night of 16 March 2021. The deepest images result in limiting magnitudes $g > 24.7$ and $r > 24.1$ (at the $3\sigma$ level). Due to the crowded nature of the field, we employed PSF photometry. The resulting residual image is presented in Fig.~\ref{fig:psffinders}, where signal above the background level at the expected position of J1539 is detected. By including this source in the PSF fit we obtain quiescent magnitudes of $g = 22.40 \pm 0.06$ and $r = 21.19 \pm 0.06$.

J1539 was also observed by the DECaPS2 survey in the $g$ and $r$-bands in 2018. Although these observations had a seeing of >1.3 arcsec, PSF photometry on the best seeing frames yield $g = 22.45 \pm 0.08$ and $r = 21.6 \pm 0.2$, consistent with our ULTRACAM quiescent magnitudes. The source was re-observed in May 2019 in the $i$ and $z$ filters under better seeing conditions (<0.7 arcsec, see Fig.~\ref{fig:decaps_finders}), with photometry yielding $i = 21.3 \pm 0.1$ and $z = 20.2 \pm 0.07$.

\citet{krimm2011} estimated a $N_{\mathrm{H}}$ value of $0.35 \times 10^{22}$\,cm$^{-2}$ (with not reported uncertainties) by fitting joint RXTE/Swift spectra. From this, we estimate $E(B-V)= 0.51 \pm 0.02$\,and derive an intrinsic color $(g-r)_0 = 0.70 \pm 0.12$, suggesting a companion star later than K0 V \citep{covey2007}. Note that this color is strictly simultaneous and, thus, representative of the true intrinsic color of the source in quiescence (i.e. free from orbital and secular variability). Our photometry indicates an outburst amplitude of $\Delta g > 4.20 \pm 0.08$. This is a lower limit since the peak outburst magnitude was missed. Therefore, we constrain the orbital period to $P_{\mathrm{orb}} \leq 21.5$\,h by using the empirical $P_{\mathrm{orb}} - \Delta V$ correlation. This results in a mean density for the companion star of > 0.24\,gr cm$^{-3}$. For a main sequence companion, this limit yields a spectral type later than A0 V. Conversely, color information constrained the spectral type to be later than K0 V, leading to a tighter upper limit on the orbital period of $P_{\mathrm{orb}} \leq 6.7$\,h, assuming that the companion is a Roche-lobe filling main sequence star.

\subsection{XTE J1650-500}
\label{sec:xtej1650}
XTE J1650-500 (hereafter J1650) was discovered by the RXTE ASM in September 2001 \citep{remillard2001} and evolved through the canonical X-ray spectral states typical of BH XRTs \citep[e.g.][]{corbel2004}. Its outburst optical counterpart was identified with a $\sim 17$\,mag object with a wide blue-band filter by \citet{castro2001}, fading to $V\sim 24$, $R\sim22$ in 2002 August when the source was found to be in quiescence \citep{garcia2002}. Recently, \citet{casares2025} have improved on these estimates by measuring a quiescent magnitude of $r=22.10\pm0.12$ from PSF photometry of DECAPS2 images (see their Appendix A). The BH nature of the compact object was dynamically confirmed by \citet{orosz2004} through a combined analysis of time-resolved FORS2 spectroscopy, taken on 10 June 2002 when J1650 was 1.4 mag brighter than in quiescence, and photometry obtained in true quiescence.

The most accurate coordinates for J1650 are from Chandra X-ray observatory observations with a positional uncertainty of 0.31 arcsec \citep{gallo2008}. We improve the location of J1650 by measuring the astrometric position of the $R=20.67 \pm 0.04$ near-quiescent optical counterpart in the FORS2 acquisition image obtaining: RA(J2000)=16:50:00.95 and DEC(J2000)=$-$49:57:44.25, with an rms uncertainty of 0.05 arcsec.

\subsection{XTE J1726-476 (=IGR J17269-4737)}
XTE J1726-476 (hereafter J1726) is an X-ray transient discovered in October 2005 by the RXTE ASM and INTEGRAL \citep{levine2005j1726, turler2005}. Subsequent outburst observations facilitated the identification of a NIR \citep[$K_s = 18.05$;][]{steeghs2005} and optical counterpart \citep[$i = 16.97 \pm 0.11$;][]{maitra2005}. \citet{lopez2019} reported a potential quiescent counterpart near the outburst optical position with $J = 21.0 \pm 0.3$ and a limiting magnitude $K_s > 17.9$, both in the AB system. Revision of the NIR images presented in \citet{lopez2019} shows that the NIR source is fully consistent with the more accurate position reported in \citet{steeghs2005} and thereby is the actual quiescent NIR counterpart.

We conducted observations of J1726 over one night in 2021 and two nights in 2023. The optical quiescent counterpart is detected in our deepest images (see Fig.~\ref{fig:finders}) at coordinates RA(J2000)=17:26:49.31 and Dec(J2000)=$-$47:38:25.63, with an rms uncertainty of 0.05 arcsec. We performed aperture photometry on this object, obtaining magnitudes of $g = 24.2 \pm 0.3$ and $r = 23.7 \pm 0.3$. For completeness, we also examined $g$ and $r$-bands images of J1726 in DECaPS2 but could not detect the object. By selecting the best seeing images, taken in 2017, we derive magnitude limits $g > 23.6$ and $r > 22.9$. These lower limits are consistent with our ULTRACAM detection. In contrast, the source is detected at longer wavelengths in May 2017 with DECaPS2 with $i = 22.3 \pm 0.1$ and $z = 21.7 \pm 0.2$ (see Fig.~\ref{fig:decaps_finders}), as well as in April 2019 with identical brightness. 

The interstellar extinction along the line of sight to J1726 is characterized by an $E(B - V)$ value of 0.428 \citep{schlafly2011}. Applying the extinction correction to the observed magnitudes, we calculate the intrinsic colors $(g-r)_0 = 0.07$ and $(i-z)_0 = 0.42$. These values are consistent with main-sequence spectral types later than F0 and M0, respectively \citep{covey2007}. However, given the large photometric errors we decided not to give constraints on the spectral type and orbital period based on the colors. On the other hand in the absence of an outburst $g$-band measurement, we can use the outburst $i$-band magnitude as a limiting factor given that optical colors in XRTs during outburst are typically blue, with $(g-i)_0 < 0$\,\citep{russell2011, buxton2012, saikia2022}. Using $E(B-V) = 0.428$\,, this implies $(g-i) < 0.68$, that combined with the observed outburst magnitude $i = 16.97 \pm 0.11$ yields $g < 17.66$ in outburst. By comparing this with our ULTRACAM quiescent detection, we derive a lower limit on the outburst amplitude of $\Delta g > 6.54$, which in turn implies an orbital period of $P_{\mathrm{orb}} \leq 10.6$\,h. Furthermore, we can constrain the mean density of a main sequence star ($> 0.98$\,gr cm$^{-3}$), which suggests a companion spectral type later than F8 V.

\begin{figure*}
	\centering \includegraphics[scale=0.5]{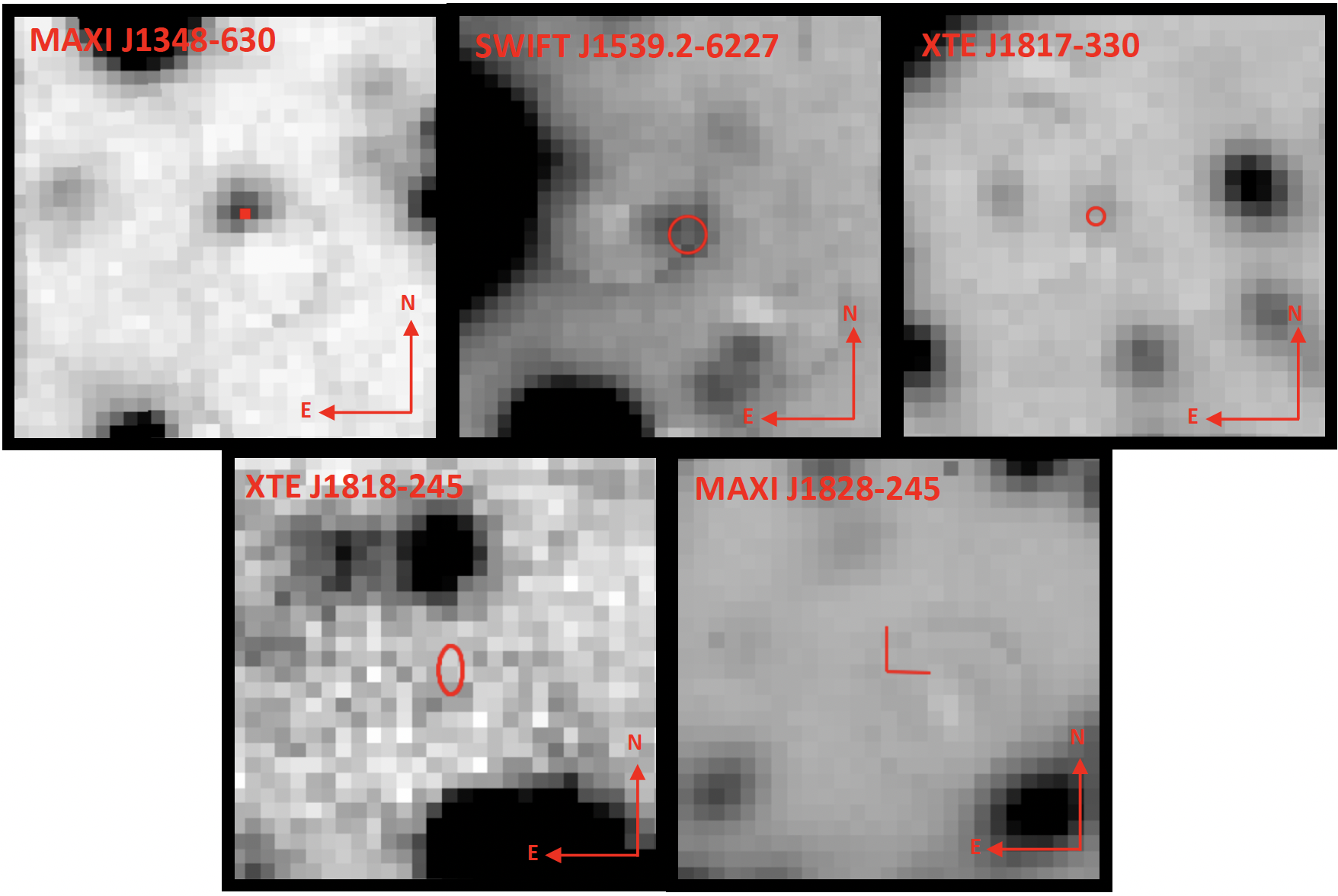}
    \caption{PSF residual $r$-band images. Object locations are indicated by red ellipses, except for J1828, which is marked with red tick marks due to its small positional error. The field of view is 10 x 10 arcsec.}
    \label{fig:psffinders}
\end{figure*}

\subsection{4U 1755-338 (= V4134 Sgr)}
V4134 Sgr is the optical counterpart to the X‐ray binary 4U 1755–338, first identified as an X‐ray source in 1972 \citep{giacconi1972}. This system is classified as quasi‐persistent, having exhibited a prolonged outburst that lasted 25 years until 1996 \citep{roberts1996}, with a subsequent reactivation reported in 2020 \citep{mereminskiy2020}. During its first active state, 4U 1755-338 reached an outburst magnitude of $V = 18.5$\,\citep{mason1985}, whereas in quiescence it faded beyond detection, with limits indicating $V > 21.96$\,\citep{wachter1998}. \citet{white1984} suggested an orbital period of 4.4 h and a high inclination system based on the presence of periodic X-ray dips. The X-ray source displays a soft spectrum \citep{white1984} and large-scale fossil jets \citep{angelini2003, kaaret2006} suggesting a BH candidate. 

We observed 4U 1755-338 on the night of 16 March 2023 and found that the system was still active. Given the clear detection of the source in our images, we took the opportunity of refining the existing astrometric coordinates. Astrometric measurements yield RA(J2000)=17:58:40.01 and Dec(J2000)=$-$33:48:28.8, with an rms uncertainty of 0.04 arcsec. These values are consistent but more precise than the latest reported position, based on observations with the e-ROSITA instrument on board the Spectrum Roentgen Gamma satellite \citep{waddell2026}. We performed differential aperture photometry on the individual exposures, and the resulting $g$- and $r$-band light curves did not show clear variability with the limited orbital coverage of our observations. We measure mean magnitudes $g = 18.91 \pm 0.04$ and $r = 18.26 \pm 0.03$, where the quoted uncertainties represent the rms variability. 

4U 1755–338 was observed by DECaPS2 during quiescence in 2017, 2018 and 2019, but the source was not detected in any band or epoch. The best-seeing images, obtained in April 2017 (see Fig.~\ref{fig:decaps_finders}), yield limiting magnitudes of $g > 22.1$, $r > 21.1$, $i > 20.6$ and $z > 20.2$. Considering the reported outburst $g$ magnitude and the proposed orbital period of 4.4 h, the expected quiescent magnitude is $g \sim 28$, well bellow our detection limits. On the other hand, the 4.4 h period implies a mean stellar density $\sim 5.7$\,gr cm$^{-3}$ that leads to an M1 V donor star.

\subsection{MAXI J1803-298}
MAXI J1803–298 (hereafter J1803) was first detected by the MAXI/GSC on May 2021 \citep{serino2021}. A likely outburst peak $g$-band magnitude of $16.06 \pm 0.01$\, was measured by \citet{matasanchez2022} \citep[see also][]{saikia2021}. Precise astrometric coordinates were subsequently provided by VLBA observations \citep{wood2023}. The system exhibited periodic absorption dips in its X-ray light curve, indicative of an orbital period of $7.02 \pm 0.18$\,h \citep{jana2022}. Based on H$\alpha$ scaling relations from spectra obtained when the system was close to quiescence, \citet{matasanchez2022} constrained the radial velocity semi-amplitude of the companion star to be between 460 and 570\,km s$^{-1}$. This constraint along with the orbital period implies a $>3$\,M$_\odot$ compact object confirming its BH nature. On the other hand, the 7.02 h orbital period implies a mean density $\sim 2.2$\,gr cm$^{-3}$ and thus a Roche-lobe filling main sequence donor with spectral type K0 V. 

J1803 was only observed on the night of 23 March 2023 under poor seeing conditions. The deepest images were obtained after co-adding all the individual data (see Fig.~\ref{fig:finders}), which result in magnitude limits of $g > 20.9$ and $r > 20.2$. The field around the position of J1803 appears very crowded with significant source confusion. In an attempt to identify the object, we performed a PSF analysis on combined DECaPS2 images under $\sim$ 0.9 arcsec with limiting magnitudes $g > 21.2$\,, $r > 20.5$\,, $i > 20.3$\,and $z > 20.1$ (Fig.~\ref{fig:decaps_finders}). After subtracting the PSF model we identify residual emission at RA(J2000)=18:03:02.80 and Dec(J2000)=$-$29:49:49.64, with an rms uncertainty of 0.04 arcsec. The astrometric coordinates are offset with respect to the exquisite VLBA position by 5$\sigma$\,in declination. This implies that the potential optical source, for which we were unable to obtain its brightness using PSF photometry, must be an interloper if not an artifact from the PSF model subtraction.

For a 7.0 h period the $\Delta V - P_{\mathrm{orb}}$ relation predicts a quiescent magnitude $g = 24.2$\,, which is fainter than our limiting magnitudes. New deeper observations under very good seeing conditions will be required to solve the optical counterpart.

\begin{table*}
	\centering
	\caption{Photometric results.}
	\label{tab:photometricresults}
	\begin{tabular}{cccccc}
		\hline
        Object & Outburst magnitudes & \multicolumn{4}{c}{Quiescent magnitudes} \\
        \cline{3-6}
        & & $g$ & $r$ & $i$ & $z$ \\ \hline
        MAXI J0637-430 & $g = 16.23 \pm 0.01$ & $> 25.8$ & $> 25.3$ & $--$ & $--$ \\
        MAXI J1348-630 & $g = 16.09 \pm 0.04$ & $23.6 \pm 0.1$ & $21.18 \pm 0.03$ & $20.5 \pm 0.1$ & $19.41 \pm 0.07$ \\
        SWIFT J1539.2-627 & $g < 18.25 \pm 0.01$ & $22.40 \pm 0.06$ & $21.19 \pm 0.06$ & $21.3 \pm 0.1$ & $20.2 \pm 0.07$ \\
        XTE J1726-476 & $g < 17.66$ & $24.2 \pm 0.3$ & $23.7 \pm 0.3$ & $22.3 \pm 0.1$ & $21.7 \pm 0.2$ \\
        4U 1755-338 & $g < 18.91 \pm 0.04$ & $> 22.1$ & $> 21.1$ & $> 20.6$ & $> 20.2$ \\
        MAXI J1803-298 & $g = 16.06 \pm 0.01$ & $> 21.2$ & $> 20.5$ & $> 20.3$ & $> 20.1$ \\
        XTE J1817-330 & $g < 14.93 \pm 0.05$ & $22.33 \pm 0.09$ & $21.24 \pm 0.07$ & $20.81 \pm 0.05$ & $20.59 \pm 0.05$ \\
        XTE J1818-245 & $V < 17.42 \pm 0.01$ & $> 21.7$ & $> 21.3$ & $> 21.1$ & $> 20.7$ \\
        MAXI J1828-249 & $g = 17.2 \pm 0.1$ & $> 22.3$ & $> 21.7$ & $> 21.7$ & $> 20.6$ \\ \hline
	\end{tabular}

\end{table*}

\subsection{XTE J1817-330}
XTE J1817–330 (hereafter J1817) was discovered on January 2006 by ASM on the RTXE \citep{remillard2006}. The ASM observations suggested a very soft X-ray source, making J1817 a BH candidate. We performed astrometry on the $g$-band outburst LDSS-3 images taken in Jannuary 2006 and presented in \citet{torres2006} to obtain the position of J1817: RA(J2000)=18:17:43.51 and DEC(J2000)=$-$33:01:07.68, with an rms uncertainty of 0.04 arcsec. These coordinates improved the radio position reported by \citet{rupen2006}. We also confirm the $g = 14.93 \pm 0.05$ outburst magnitude reported in \citet{torres2006} by calibrating the field with DECaPS2.

\begin{figure*}
	\centering \includegraphics[width=\textwidth,height=\textheight,keepaspectratio]{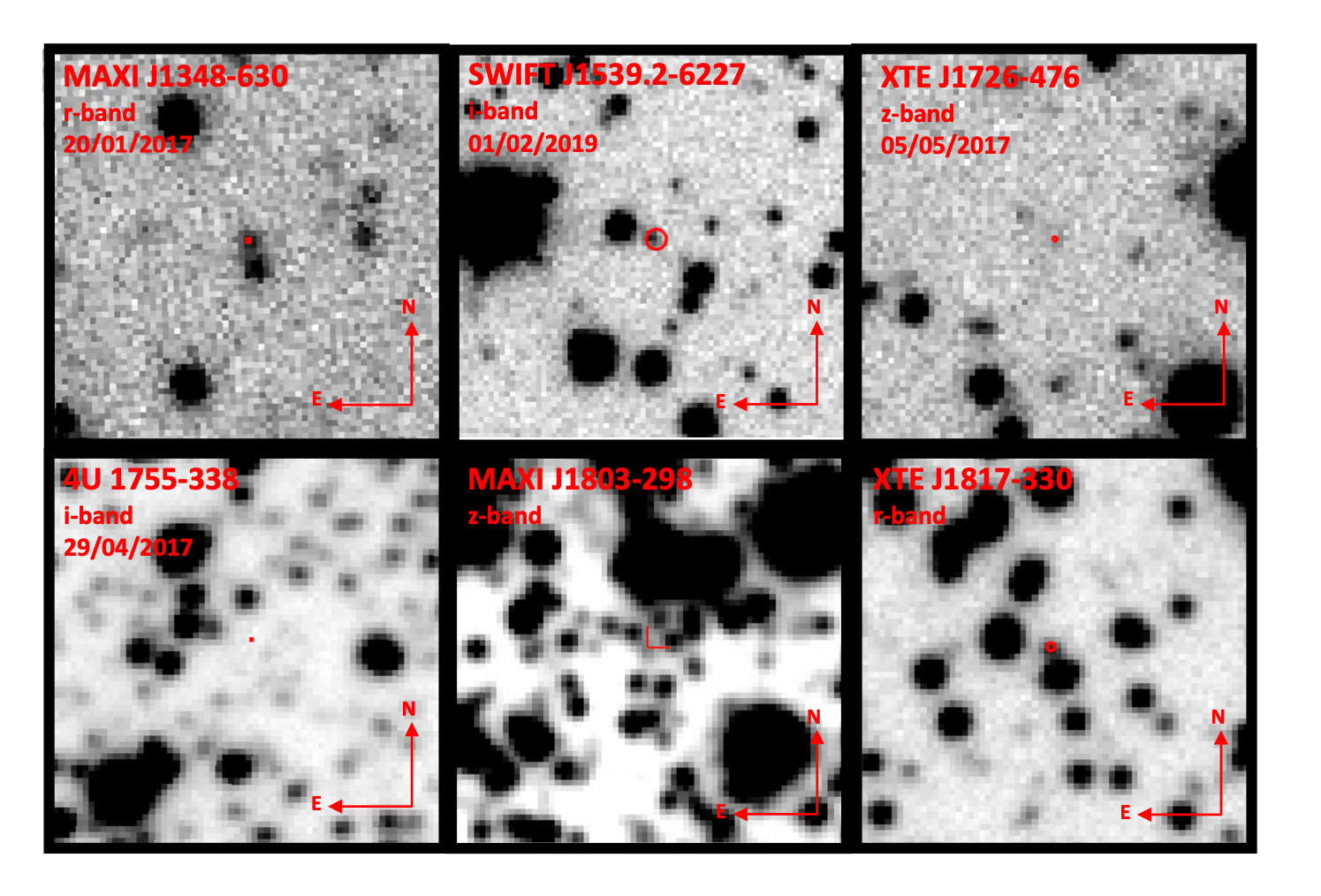}
    \caption{Best identifications from DECaPS2 images, with each field of view measuring 20 × 20 arcsec. The positions of the objects are marked with red ellipses, accounting for the positional uncertainties reported in the literature (see Table~\ref{tab:coords}), except for 4U 1755-338, where we plot our refined coordinates based on the outburst counterpart detected in the NTT images. For the sake of visibility, we mark the radio position of J1803 with red tick marks.}
    \label{fig:decaps_finders}
\end{figure*}

J1817 ULTRACAM photometry was taken on the night of March 16, 2021. The deepest $r$-band image is shown in Fig.~\ref{fig:finders} for which we obtain $3\sigma$ limiting magnitudes of $g > 22.2$ and $r > 21.5$. Given the high level of crowding, we proceeded to perform PSF photometry. The resulting residual $r$-band image is displayed in Fig.~\ref{fig:psffinders}, where an underlying residual emission is detected at the position of J1817. However, we were unable to obtain the brightness of the potential source with PSF photometry. We also examined DECaPS2 images of the field taken in May 2018 (see Fig.~\ref{fig:decaps_finders}). In this case, J1817 was detected and identified with the source with ID = 4809228675506884372 with fluxes which correspond to magnitudes $g = 22.33 \pm 0.09$, $r = 21.24 \pm 0.07$, $i = 20.81 \pm 0.05$ and $z = 20.59 \pm 0.05$. These magnitudes are consistent with our own PSF photometry of the DECaPS2 images.

\citet{sala2007} estimated $N_{\mathrm{H}} = (1.52 \pm 0.05) \times 10^{21}$\,cm$^{-2}$ based on XMM-Newton and INTEGRAL observations. We used this to obtain the $E(B-V) = 0.22 \pm 0.02$\,and derive the intrinsic colors $(g-r)_0=0.86 \pm 0.14$ and $(r-i)_0 = 0.30 \pm 0.11$, with both mean values suggesting a K3 V companion star. On the other hand, from the quiescent DECaPS2 $g$-band detection, we derive an outburst amplitude $\Delta g > 7.4$ which, combined with the empirical $P_{\mathrm{orb}} - \Delta V$ correlation, imply an orbital period of $P_{\mathrm{orb}} \leq 8.2$\,h. The mean stellar density for that period ($> 1.6$\,gr cm$^{-3}$) leads to a > G6 main sequence star, which is not very restrictive. Given the consistency in the spectral type derived from the two DECaPS2 colors we adopt a K3 V companion star for this object. Assuming a Roche-lobe filling main sequence star this spectral type implies an orbital period of 6.6 h.

\subsection{XTE J1818-245}
The X-ray transient XTE J1818-245 (hereafter J1818) was initially detected on August 2005 by the ASM on RTXE \citep{levine2005} showing a soft X-ray spectrum typical of BH XRTs. The optical counterpart was identified with $V = 17.42 \pm 0.01$ a few days after the X-ray discovery \citep{steeghs2005j1818}. A NIR counterpart was also detected with an apparent magnitude of $K_s = 16.18 \pm 0.02$ during outburst and $K_s = 19.99 \pm 0.22$ in quiescence \citep{lopez2019}, both in the AB system. We derived the position of J1818 from the astrometrically calibrated NIR outburst image presented in \citet{lopez2019}: RA(J2000)=18:18:24.43 and DEC(J2000) = -24:32:18.08, with an rms uncertainty of 0.08 arcsec. These coordinates improved the radio position reported by \citet{rupen2005}.

J1818 was observed on the night of 16 March 2021. The deepest images reached limiting magnitudes $g > 21.7$ and $r > 20.8$. Since the source is located in a crowded field, PSF photometry was performed in both $g$ and $r$-bands. The resulting $r$-band residual, displayed in Fig.~\ref{fig:psffinders}, shows no detectable signal at the position of J1818. Likewise, our search using Pan-STARRS images and PSF photometry on the frames with best seeing yielded no detections, with limiting magnitudes of $g > 21.7$, $r > 21.3$, $i > 21.1$ and $z > 20.7$.

By applying our deepest quiescent limit along with the outburst magnitude reported by \citet{steeghs2005}, we find $\Delta g > 4.3$, which implies $P_{\mathrm{orb}} \leq 20.8$\,h. This limit on the orbital period corresponds to a mean stellar density $> 0.25$\,gr cm$^{-3}$, suggesting that the companion star is later than A0 V.

\subsection{MAXI J1828-249}
MAXI J1828-249 (hereafter J1828) was discovered on October 2013 by the MAXI telescope showing a soft X-ray spectrum reminiscent of a BH XRT \citep{nakahira2013}. The optical counterpart was detected during outburst with $g = 17.2 \pm 0.1$ \citep{arne2013}. We adopt this as the peak magnitude since other contemporaneous observations show no major decrease in flux in the early phase of the outburst \citep{avanzo2013, grebenev2016}. A precise astrometric position of the source was given by \citet{kennea2013}. \citet{lopez2019} reported a quiescent NIR magnitude of $K_S = 20.82 \pm 0.09$ in the AB system. Subsequent outbursts in 2015 and 2016 were detected at mid-infrared and optical wavelengths \citep[e.g.][]{mowlavi2024, chris2024} and Gaia observations provided a refined astrometric position for the transient (see Table~\ref{tab:coords}).

J1828 was observed during the night of 23 March 2023. The deepest images yield limiting magnitudes $g > 22.3$ and $r > 21.7$. Given that there is a bright star separated by 1.6 arcsec from the expected position of J1828 (see Fig.~\ref{fig:finders}), we performed PSF photometry and studied the residual after subtracting this source. The residual $r$-band image is displayed in Fig.~\ref{fig:psffinders}, but no star is detected at the position of J1828. On the other hand, we examined the individual Pan-STARRS images but no detection was found. The best-seeing frames deliver limiting magnitudes of $g > 21.7$, $r > 21.6$, $i > 21.7$ and $z > 20.6$. From the outburst amplitude, and using the $P_{\mathrm{orb}} - \Delta V$ correlation we find $P_{\mathrm{orb}} \leq 15.4$\,h. This limit implies a mean stellar density for the companion star of $\geq 0.5$\,gr cm$^{-3}$\,, which suggests a spectral type later than A6 V for the unseen companion. 

\begin{table*}
	\centering
	\caption{Binary constraints assuming main sequence companions.}
	\label{tab:binaryconstraints}
	\begin{tabular}{ccc|cc}
		\hline
        Object & \multicolumn{2}{c}{Orbital period (h) from} & \multicolumn{2}{c}{Spectral type from} \\
        \cline{2-3} \cline{4-5}
         & outburst amplitude & quiescent colors & $P_{\mathrm{orb}}$\,in column \#2 & quiescent colors \\ \hline
        MAXI J0637-430 & $\leq 4.5$ & $--$ & $> \mathrm{M1\,V}$ & $--$ \\
        MAXI J1348-630 & $\leq 7.9$ & $\leq 10.1$ & $> \mathrm{G8\,V}$ & $> \mathrm{K2\,V}$ \\
        SWIFT J1539.2-627 & $\leq 21.5$ & $\leq 6.7$ & $> \mathrm{A0\,V}$ & $> \mathrm{K0\,V}$ \\
        XTE J1726-476 & $\leq 10.6$ & $--$ & $> \mathrm{F8\,V}$ & $--$ \\
        4U 1755-338 & \multicolumn{2}{c|}{$4.4^{a}$} & \multicolumn{2}{c}{$\approx \mathrm{M1\,V}^{b}$} \\
        MAXI J1803-298 & \multicolumn{2}{c|}{$7.02 \pm 0.18^{c}$} & \multicolumn{2}{c}{$\approx$\,K0 V$^{d}$} \\
        XTE J1817-330 & $\leq 8.2$ & $\approx 6.6$ & $> \mathrm{G6\,V}$ & $\approx \mathrm{K3\,V}$ \\
        XTE J1818-245 & $\leq 20.8$ & $--$ & $> \mathrm{A0\,V}$ & $--$ \\
        MAXI J1828-249 & $\leq 15.4$ & $--$ & $> \mathrm{A6\,V}$ & $--$ \\ \hline
	\end{tabular}
    \tablefoot{
    $^{a}$ \citet{wachter1998}; $^{b}$ Spectral type for a Roche lobe filling main sequence with $P_{\mathrm{orb}} = 4.4$\,h; $^{c}$ Measured from X-ray dips \citep{jana2022}; $^{d}$ Spectral type for a Roche lobe filling main sequence with $P_{\mathrm{orb}} = 7.02 \pm 0.18$\,h.}
\end{table*}

\section{Discussion}
\label{final_considerations}
The constraints on binary physical parameters derived in this paper should be treated with caution as they are subject to several caveats. To start with, the intrinsic color of the quiescent counterparts can sometimes be dominated by systematic errors introduced by the chosen extinction law and the determination of the color excess \citep[see ][for thorough review]{hynes2005}. We have adopted the average extinction for the Galaxy $A_V/E(B-V) = 3.1$\,, and therefore we are prone to errors due to the uncertain extinction curve in the direction of the targets.

For the determination of the color excess from the $N_H$ derived from X-ray absorption, we have adopted an average $N_H/A_V$ ratio of $(2.21 \pm 0.09) \times 10^{21}$ for the galaxy that is obtained assuming interstellar solar abundances \citep{guver2009}. However, we note that the $N_H/A_V$ ratio can be $\sim 20$ per cent larger when using subsolar abundances \citep{zhu2017}. This, together with the possibility of having absorption local to the source, implies that we could be overestimating the value of $E(B-V)$, thus obtaining an upper limit to the color excess.

When available, we have used $N_H$ measurements from the literature, assuming the formal errors from X-ray fitting models. For example, in the case of XTE J1817-330 the formal $N_H$ error ($1.52 \pm 0.05 \times 10^{21}$) translates into a very small uncertainty of $\pm 0.05$ mags in the quiescent color. Consequently, the spectral classification could be dominated by systematic uncertainties rather than by the photometric errors of the quiescent magnitudes.

In the case of MAXI J1348-630, however, different authors have reported a wide range of $N_H$ values depending on the X-ray state and the fitted model. As noted by \citet{zhang2022} these values most likely reflect a degeneracy in the fitted parameters rather than real physical changes. In this particular case, we have adopted $8.6 \times 10^{21}$ as the most stable value near the peak of the outburst, as favored by independent analyses of \citet{tominaga2020, zhang2022, liu2022, dai2023}. Allowing for a conservative uncertainty of $\pm 2.5 \times 10^{21}$ to accommodate the range of $N_H$ values measured in all the observations performed during the hard/soft intermediate state and soft state \citep{zhang2022}, this would translate into a systematic error of $\pm 1.4$ mag in the quiescent color. Therefore, in this particular case the spectral type implied by the quiescent color is essentially unrestricted, although it is ultimately constrained by the upper limit in orbital period derived from the amplitude of the outburst.

Only for XTE J1726-476 we have adopted an upper limit to $E(B-V)$ (thereby an upper bound to the spectral type) from the integrated color excess over the galactic sight line. Here, the angular resolution of the dust map is a potential source of systematic uncertainties due to the inability to trace the dust distribution of the local gas in the direction to the source. Additionally, as noted in Sect.~\ref{sec:extinctionlaws}, any spectral type constraint based on quiescent colors should be regarded as a mere upper limit since the potential contribution of an accretion disc has been neglected.

Regarding the orbital period constraints, it is important to acknowledge that the exploited $\Delta V - P_{orb}$ relation was obtained from just eight systems covering orbital periods in the range $\sim 5-24$\,h \citep{shahbaz1998}. \citet{corral2018} have noted that the correlation breaks down at short orbital periods $\leq 5$\,h. Also, this relation is expected to overestimate the outburst amplitude for high inclination systems because of projection effects \citep{miller2011, kuulkers2013}. consequently, the orbital periods derived in this work should also be treated as merely orientative rather than precise measurements, as they were obtained in a best effort given the information available. These orbital period constraints, together with the apparent magnitudes, should serve to design new observations aimed at establishing (or better constraint) the XRT orbital period and companion spectral type.

In four out of five targets (4U 1755-338, MAXI J1803-298, XTE J1818-245, MAXI J1828-249) the non detection is mostly driven by poor observing conditions, 
sometimes also coupled with large interstellar extinction.  Only in the case of MAXI J0637-430 the deep magnitude limit at $r > 25.3$ is caused by the
intrinsic faintness of the object (as implied by the large outburst  amplitude) and perhaps a large distance. On the other hand, given our predicted magnitude of $g \sim 28$ for its quiescent counterpart, we would not have expected to detect 4U 1755-338, even under ideal observing conditions. Follow-up observations with new generation survey telescopes such as LSST are expected to provide first detections on these five targets or more stringent limits, from which further physical constraints will be established.

Up to now we have adopted the conservative approach of deriving binary constraints assuming that the companion stars in our nine XRTs are Roche-lobe filling main sequence stars. We note, however, that for $P_{\mathrm{orb}} \geq 15$\,h the main sequence companions would be intermediate-mass stars with $M_2 \geq 1.5$\,M$_\odot$ and spectral types earlier than $\sim$F1. Only three BH XRTs with intermediate-mass donors are known (V4641 Sgr, GRO J1655-40 and 4U 1543-475) and they all have $P_{\mathrm{orb}} > 1$\,d and small outburst amplitudes of $\lesssim 4$\,mag\footnote{Only V4641 Sgr has shown an outburst amplitude of 5.3 mag during a brief super-Eddington X-ray flare in 1999 \citep{kato1999}. The amplitude of the other eight recorded outbursts of V4641 Sgr has always been $< 4$\,mag.}. Conversely, the nine XRTs studied in this work have either large outburst amplitudes and/or short orbital periods, a strong indication that the companions are low-mass stars. 

Recently, \citet{casares2025} have derived an empirical correlation between $T_{\mathrm{eff}}$ and orbital period from a sample of 17 BH XRTs with low-mass donors. Sixteen systems trace a very narrow track in the $P_{\mathrm{orb}} - T_{\mathrm{eff}}$ plane, an indication that they all follow the same evolutionary path. Only XTE J1118+480 appears 500 K hotter than expected, suggesting that its donor has descended from an intermediate-mass star that was significantly evolved before the onset of mass transfer, and this has been confirmed through evidence of processed CNO material \citep{haswell2002}. Under the hypothesis that the companions in our nine XRTs are low-mass stars we have applied the empirical B2 equation from \citet{casares2025} to our $P_{\mathrm{orb}}$ constraints, together with the $T_{\mathrm{eff}}$-spectral type scale of \citet{pecaut2013}, and derived new spectral type constraints that we think are more reliable than those provided in Table~\ref{tab:binaryconstraints}. Our final system constraints are summarized in Table~\ref{tab:finalconstraints}. In any case, we warn that even these are tentative, since one cannot rule out the possibility that the donor star in a particular system has evolved from an intermediate-mass star, in which case the spectral type constraint would be biased low.

\begin{table}
	\centering
	\caption{Final spectral types based on the $P_{\mathrm{orb}}-T_{\mathrm{eff}}$ relation.}
	\label{tab:finalconstraints}
	\begin{tabular}{ccc}
		\hline
        Object & \multicolumn{2}{c}{Spectral type from $P_{\mathrm{orb}}$ based on}\\
        \cline{2-3}
        & outburst amplitude & quiescent colors \\ \hline 
        MAXI J0637-430 &
            \begin{tabular}{c} > M3 V \end{tabular} &
	        \begin{tabular}{c} $--$ \end{tabular} \\ \hline
        MAXI J1348-630 &
            \begin{tabular}{c} > K4 V \end{tabular} &
	        \begin{tabular}{c} > K4 V \end{tabular} \\ \hline
		SWIFT J1539.2-6227 &
            \begin{tabular}{c} > K4 V \end{tabular} &
	        \begin{tabular}{c} > K6 V \end{tabular} \\ \hline
        XTE J1726-476 &
            \begin{tabular}{c} > K4 V \end{tabular} &
	        \begin{tabular}{c} $--$ \end{tabular} \\ \hline   
        4U 1755-338 &
            \multicolumn{2}{c}{$\sim$\,M3 V}  \\ \hline     
        MAXI J1803-298 &
            \multicolumn{2}{c}{$\sim$\,K5 V}  \\ \hline  
        XTE J1817-330 &
            \begin{tabular}{c} > K4 V \end{tabular} &
	        \begin{tabular}{c} $\approx$\,K5 V \end{tabular} \\ \hline    
        XTE J1818-245 &
            \begin{tabular}{c} > K3 V \end{tabular} &
	        \begin{tabular}{c} $--$ \end{tabular} \\ \hline
        MAXI J1828-249 &
            \begin{tabular}{c} > K3 V \end{tabular} &
	        \begin{tabular}{c} $--$ \end{tabular} \\ \hline     
	\end{tabular}

\end{table}

\section{Conclusions}
In this work, we present an optical photometric campaign of nine BH XRTs that currently have non established quiescent optical counterparts. Our observations successfully identify counterparts for four targets, while for the remaining five sources, we obtain $3 \sigma$\,lower limits for their magnitudes in quiescence. Additionally, one system (4U 1755-338) was found in outburst. Complementary photometry, obtained during outburst, is also collected and reviewed. Based on the outburst amplitudes and quiescent colors, we derive constraints on the orbital periods and spectral types of the companion stars, assuming a Roche-lobe filling main sequence star scenario (Table~\ref{tab:binaryconstraints}). Finally, by adopting the empirical $P_{\mathrm{orb}}-T_{\mathrm{eff}}$\,relation from equation B2 of \citet{casares2025} for low-mass companions in BH XRTs we obtain more realistic spectral type constraints that are summarized in Table~\ref{tab:finalconstraints}. 

Objects detected with $r < 22$ magnitudes are prime candidates for future dynamical studies through time-resolved optical spectroscopy, which will lead to confirmed orbital periods and BH mass determinations. Dynamical constraints on fainter counterparts will require the use of future $> 30$\,m class telescopes or H$\alpha$ scaling relations \citep[e.g. ][]{casares2015, casares2016, casares2022}.

\begin{acknowledgements}
We thank the anonymous referee for providing useful and constructive suggestions. JC and MAPT acknowledge support by the Spanish Ministry of Science via the Plan de Generaci\'on de Conocimiento through grants PID2022-143331NB-100 and PID2021-124879NB-I00, respectively. SNU is supported by the FPI grant PREP2022-000508, also under program PID2022-143331NB-100. VSD and ULTRACAM are supported by the UK Science and Technology Facilities Council (STFC). M.A.P. acknowledges support through the Ramón y Cajal grant RYC2022-035388-I, funded by MCIU/AEI/10.13039/501100011033 and FSE+. PGJ is funded by the European Union (ERC, Starstruck, 101095973). Views and opinions expressed are however those of the authors only and do not necessarily reflect those of the European Union or the European Research Council Executive Agency. Neither the European Union nor the granting authority can be held responsible for them. 

\end{acknowledgements}

\bibliographystyle{aa} 
\bibliography{references} 

\appendix

\begin{table*}
	\centering
    \section{Tables}
	\caption{Log of the NTT observations.}
	\label{tab:log}
	\begin{tabular}{ccccccc}
		\hline
            Object & BlackCAT ID & Date & \shortstack{Total\\ exposure (h)} & \shortstack{Seeing\\ (arcsec)} & \shortstack{Seeing cut-off\\ (arcsec)} &  \shortstack{Effective\\ exposure (h)} \\ \hline 
        MAXI J0637-430 &
	    \begin{tabular}{c} 68
            \end{tabular} &
        \begin{tabular}{c} $15/03/2021$ \\
        $16/03/2021$\\
        $23/03/2023$\\
        $24/03/2023$\\
            \end{tabular} & 
        \begin{tabular}{c} $1.71$ \\
        1.70 \\
        1.01 \\
        0.36 \\
            \end{tabular} &
        \begin{tabular}{c} 1.4 - 2.6 \\
        0.9 - 1.5 \\
        0.8 - 1.2 \\
        1.7 - 2.8 \\
            \end{tabular} &
        \begin{tabular}{c} 1.00 \\
            \end{tabular} &
        \begin{tabular}{c} 2.1 \\
            \end{tabular} \\ \hline 
        MAXI J1348-630 &
	    \begin{tabular}{c} 67 \\
            \end{tabular} & 
        \begin{tabular}{c} $15/03/2021$ \\
        $16/03/2021$ \\
        $23/03/2023$\\
            \end{tabular} & 
        \begin{tabular}{c} 2.87\\
        2.74 \\
        4.09 \\
            \end{tabular} &
        \begin{tabular}{c} 1.6 - 2.3 \\
         1.0 - 2.1 \\
         0.9 - 1.5 \\
            \end{tabular} &
        \begin{tabular}{c} 1.20 \\
            \end{tabular} &
        \begin{tabular}{c} 4.4 \\
            \end{tabular} \\ \hline
		SWIFT J1539.2-6227 & 
	    \begin{tabular}{c} 46 \\
            \end{tabular} & 
        \begin{tabular}{c} $16/03/2021$ \\
            \end{tabular} & 
        \begin{tabular}{c} 4.18 \\
            \end{tabular} &
        \begin{tabular}{c} 1.1 - 1.7 \\
            \end{tabular} &
        \begin{tabular}{c} 1.25 \\
            \end{tabular} &
        \begin{tabular}{c} 0.5 \\
            \end{tabular} \\ \hline
        XTE J1726-476 &
	    \begin{tabular}{c} 41 \\
            \end{tabular} & 
        \begin{tabular}{c} $16/03/2021$ \\
        $23/03/2023$\\
        $24/03/2023$\\
            \end{tabular} & 
        \begin{tabular}{c} 1.05\\
        0.72 \\
        2.92 \\
            \end{tabular} &
        \begin{tabular}{c} 1.1 - 1.5 \\
         1.0 - 1.6 \\
         1.2 - 2.3 \\
            \end{tabular} &
        \begin{tabular}{c} 1.30 \\
            \end{tabular} &
        \begin{tabular}{c} 1.3 \\
            \end{tabular} \\ \hline
        4U 1755-338 &
	    \begin{tabular}{c} 4 \\
            \end{tabular} & 
        \begin{tabular}{c} $23/03/2023$ \\
            \end{tabular} & 
        \begin{tabular}{c} 0.71\\
            \end{tabular} &
        \begin{tabular}{c} 0.9 - 1.3\\
            \end{tabular} &
        \begin{tabular}{c} $--$ \\
            \end{tabular} &
        \begin{tabular}{c} $--$ \\
            \end{tabular} \\ \hline  
        MAXI J1803-298 &
	    \begin{tabular}{c} 70 \\
            \end{tabular} & 
        \begin{tabular}{c} $23/03/2023$ \\
            \end{tabular} & 
        \begin{tabular}{c} 0.89\\
            \end{tabular} &
        \begin{tabular}{c} 1.2 - 2.4 \\
            \end{tabular} &
        \begin{tabular}{c} $--$ \\
            \end{tabular} &
        \begin{tabular}{c} $--$ \\
            \end{tabular} \\ \hline   
        XTE J1817-330 &
	    \begin{tabular}{c} 42 \\
            \end{tabular} & 
        \begin{tabular}{c} $16/03/2021$ \\
            \end{tabular} & 
        \begin{tabular}{c} 1.04 \\
            \end{tabular} &
        \begin{tabular}{c} 1.1 - 1.7 \\
            \end{tabular} &
        \begin{tabular}{c} 1.25 \\
            \end{tabular} &
        \begin{tabular}{c} 0.15 \\
            \end{tabular} \\ \hline     
        XTE J1818-245 &
	    \begin{tabular}{c} 40 \\
            \end{tabular} & 
        \begin{tabular}{c} $16/03/2021$ \\
            \end{tabular} & 
        \begin{tabular}{c} 1.40 \\
            \end{tabular} &
        \begin{tabular}{c} 1.1 - 1.5 \\
            \end{tabular} &
        \begin{tabular}{c} 1.25 \\
            \end{tabular} &
        \begin{tabular}{c} 0.6 \\
            \end{tabular} \\ \hline 
        MAXI J1828-249 &
	    \begin{tabular}{c} 57 \\
            \end{tabular} & 
        \begin{tabular}{c} $23/03/2023$ \\
            \end{tabular} & 
        \begin{tabular}{c} 1.00\\
            \end{tabular} &
        \begin{tabular}{c} 1.1 - 1.4 \\
            \end{tabular} &
        \begin{tabular}{c} 1.20 \\
            \end{tabular} &
        \begin{tabular}{c} 0.7 \\
            \end{tabular} \\ \hline      
	\end{tabular}
\end{table*}

\end{document}